\documentclass[preprint2]{aastex}

\slugcomment{Submitted to AJ}

\shorttitle{Metal abundances in open clusters}
\shortauthors{Carraro et al.}

\begin{document}


\title{Metal abundances in extremely distant Galactic old open clusters. \\
I. Berkeley 29 and Saurer 1$^1$}\footnotetext[1]{The data presented
herein were obtained at the W.M. Keck Observatory, which is operated
as a scientific partnership among the California Institute of
Technology, the University of California and the National Aeronautics
and Space Administration. The Observatory was made possible by the
generous financial support of the W.M. Keck Foundation}


\author{Giovanni Carraro\altaffilmark{a,b}}
\affil{Departamento de Astr\'onomia, Universidad de Chile,
    Casilla 36-D, Santiago de Chile, Chile}
\email{gcarraro@das.uchile.cl}

\author{Fabio Bresolin}
\affil{Institute for Astronomy, 2680 Woodlawn Drive, Honolulu, HI 96822, USA}
\email{bresolin@ifa.hawaii.edu}

\author{Sandro Villanova}
\affil{Dipartimento di Astronomia, Universit\`a di Padova, Vicolo Osservatorio
5, I$-$35122, Padova, Italy}
\email{villanova@pd.astro.it}

\author{Francesca Matteucci}
\affil{Dipartimento di Astronomia dell'Universit\`a di Trieste, via Tiepolo 11,
I$-$34131 Trieste, Italy}
\email{matteucci@ts.astro.it}

\author{Ferdinando Patat}
\affil{ESO, K. Schwarzschild Str. 2, 85748 Garching, Germany }
\email{fpatat@eso.org}

\and

\author{Martino Romaniello}
\affil{ESO, K. Schwarzschild Str. 2, 85748 Garching, Germany }
\email{mromanie@eso.org}


\altaffiltext{a}
{Dipartimento di Astronomia, Universit\`a di Padova, Vicolo Osservatorio 5,
I$-$35122, Padova, Italy}
\altaffiltext{b}{Astronomy Department, New Haven, CT 06520$-$8101, USA}


\begin{abstract}
We report on high resolution spectroscopy of four giant stars in the
Galactic old open clusters Berkeley~29 and Saurer~1 obtained with HIRES at the
Keck telescope.  These two clusters possess the largest galactocentric
distances insofar known for open star clusters, and therefore are
crucial objects to probe the chemical pattern and evolution of the
outskirts of the Galactic disk. We find that $[Fe/H]=-0.38\pm0.14$ and
$[Fe/H]=-0.44\pm0.18$ for Saurer 1 and Berkeley~29, respectively.
Based on these data, we first revise the fundamental parameters of
the clusters, and then discuss them in the context of the
Galactic disk radial abundance
gradients. Both clusters seem to significantly deviate from the
general trend, suggesting that the outer part of the Galactic disk
underwent a completely different evolution compared to the inner disk.
In particular Berkeley~29 is clearly associated with the Monoceros
stream, while Saurer~1 exhibits very different properties.  The
abundance ratios suggest that the chemical evolution of the outer disk
was dominated by the Galactic halo.

\end{abstract}



\keywords{open clusters: general ---
open clusters: individual (\objectname{Berkeley 29}),
open clusters: individual (\objectname{Saurer 1})}


\section{Introduction}
The detailed knowledge of the present day Galactic disk abundance gradient
and its evolution with time is one of the basic ingredients of any
chemical evolution model which aims to predict the properties of the
Galactic disk (\citealt{mat04} and references therein). 
Compared to other indicators (HII
regions, B stars, planetary nebulae and Cepheids), old open clusters
(OCs)
present the advantage of both sampling almost the entire disk and
covering basically all of the history of the disk from its infancy to
now. In recent years, renewed efforts to probe the chemical abundances
in old OCs have taken place, by means of high resolution
spectroscopy of member stars \citep{fri03,bra01, pet98, carre04}.  In
fact, the Galactic disk radial abundance gradients
\citep{fri93,car98,fri02} mostly rely  on medium resolution spectra,
and the derived metallicities turned out to be
quite different in several cases when high resolution
spectra are available (see, e.g.,  \citealt{carre04}).\\
An additional drawback stems from the fact that it is
quite common to consider the radial abundance gradient observed in the
solar vicinity as representative of the whole disk, due to the lack of
metallicity estimates in very distant OCs. It is therefore
highly desirable to obtain information on the chemical compositions of
old OCs spanning the widest possible range in galactocentric
distances, covering in particular the relatively unexplored outer disk
of the Galaxy.

In this paper we present high resolution spectra of four giant stars
in the old OCs Berkeley 29 and Saurer 1.  According to
\citet{kal94} and \citet{car03} these two objects are the most distant
OCs insofar detected in the Milky Way (beyond 19 kpcs
from the Galactic center), although
the precise distances are not yet very well known, due to
uncertainties in the reddening and metallicity.
In fact only photometric estimates of the metallicity are available
for these clusters, which are based mainly on isochrone fitting or
comparison with other oprn clusters.  \citet{kal94} reports for Berkeley~29
an age
of 4 Gyrs and and the very low metal content $[Fe/H] \approx -1.0$,
while \citet{car03} report for Saurer~1 an age of 5 Gyrs  and a metal content
$[Fe/H] \approx -0.5$
Thus, they represent 
very useful targets to probe the metal abundance of the Galactic
disk outskirts, allowing us to significantly enlarge the distance
baseline of the radial abundance gradient. Here we present new
metallicity estimates for the two clusters and derive updated
estimates of their age and distance, and discuss their role in shaping
the radial abundance gradient.\\ The layout of the paper is as
follows. Section~2 and 3 illustrate observations and reduction
strategies, while Section~4 deals with radial velocity determinations.
In Section~5 we derive the stellar abundances and in Section~6 we
revise the cluster fundamental parameters. The results of this paper
are then discussed in Section~7 and, finally, summarized in Section~8.

\section{Observations}
The observations were carried out on the night of January 14, 2004 at
the W.M. Keck Observatory under photometric conditions and typical
seeing of 1\arcsec.  The HIRES spectrograph \citep{vog94} on the Keck
I telescope was used with a 1.15 x 7 arcsec slit to provide a spectral
resolution R = 34,000 in the wavelength range 5720$--$8160~\AA\/ in 19
different orders on the 2048$\times$2048 CCD detector. A blocking
filter was used to remove second$-$order contamination from blue
wavelengths. Three exposures of 2400 seconds were obtained for the two
stars in Berkeley~29. For Saurer~1$-$91 and Saurer~1$-$122 we took two
exposures of  3000 and 2700 seconds, respectively. For the
wavelength calibration, spectra of a thorium$-$argon lamp were secured
after the set of exposures for each star was completed. The radial
velocity standard HD~26162 was observed at the beginning of the night,
together with the spectrophotometric standard G191B2B.\\ In Fig.~1 we
show a finding chart for the two clusters where the four observed
stars are indicated, while in Fig.~2 we show the position of the stars
in the Color$-$Magnitude Diagram (CMD).

\section{Data Reduction}
Images were reduced using IRAF\footnote[2]{IRAF is distributed by the
National Optical Astronomy Observatories, which are operated by the
Association of Universities for Research in Astronomy, Inc., under
cooperative agreement with the National Science Foundation.},
including bias subtraction, flat$-$field correction, extraction of
spectral orders, wavelength and flux calibration, and continuum
subtraction.  The single orders were merged into a single spectrum and
the spectra of each star were combined to remove cosmic rays.
A spectral region containing a large number
of telluric lines was instead used to correct the spectra for flexures
of the instrument and off$-$center slit pointing; the error on this
correction was about 0.05 km s$^{-1}$. An example of sprectrum, with a few
line indicated, is shown in Fig.~3.

\section{Radial Velocities} 
No radial velocity estimates are available for Berkeley~29 and Saurer~1.
The radial velocities of the target stars were measured using the IRAF
FXCOR task, which cross$-$correlates the object spectrum with the
template (HD~26162); the peak of the cross$-$correlation was fitted with a
gaussian curve after  rejecting spectral regions contaminated by
telluric lines ($\lambda > 6850$~\AA).  In order to check our
wavelength calibration we also measured the radial velocity of
HD~26162 itself, using the Doppler shifts of various spectral lines.
We obtained a
radial velocity of 24.8$\pm$0.1 km s$^{-1}$, which perfectly matches the catalogue 
value (24.8 km/sec, \citealt{wiel99}).
The final error in the radial
velocities was typically about 0.1 km s$^{-1}$.  The two stars we measured in
each clusters have compatible radial velocities (see Table~1), and
are considered, therefore,  {\it bona fide} cluster members.

\section{ABUNDANCE ANALYSIS}

\subsection{Atomic parameters and equivalent widths}
We derived equivalent widths of spectral lines by using the
interactive software {\it SUPERSPECTRE} (freely distributed by Chris
Sneden, University of Texas, Austin).  This software has the advantage
of providing a graphical visualisation of the continuum and of the
measured width.  Repeated measurements show a typical error of about
4$-$5 m\AA, also for the weakest lines. We checked the equivalent widths
derived from {\it SUPERSPECTRE} by measuring them also using standard
IRAF routine {\it SPLOT}, and
found a fair agreement, with maximum differences amounting to 5m\AA.
  The line list 
(FeI, FeII, Mg, Si, Ca, Al,, Na, Ni and Ti, see Table~3) was
taken from \citet{fri03}, who considered only lines with equivalent
widths narrower than 150m\AA, in order to avoid non$-$linear effects in the LTE
analysis of  the
spectral features. Oxygen lines were taken from
\citet{cav97}. We are aware that the use of high excitation O triplet lines is
controversial. Most problems however arise in metal poor stars ($[Fe/H] \leq 1 $)
and for temperatures larger than 6200 $^0 K$ (\citealt{kin95}). 
Our stars are metal richer
than this value and cooler, a fact which makes us confident about the
derivation of the O abundance .

\subsection{Atmospheric parameters}
Initial estimates of the atmospheric parameter $T_{eff}$ were obtained
from photometric observations in the optical (BVI) and infrared (JHK)
from 2MASS.  VI data were available for Saurer~1
\citep{car03} and BV for Berkeley~29 \citep{kal94}.  Reddening values
are E(V$-$I)= 0.18 and E(B$-$V)= 0.21, respectively. First guess effective
temperatures were derived from the (V$-$I)--$T_{eff}$ and
(B$-$V)--$T_{eff}$ relations, the former from \citet{alo99} an the
latter from \citet{gra96}.  We then adjusted the effective temperature
by minimizing the slope of the abundances obtained from Fe~I lines
with respect to the excitation potential in the curve of growth
analysis.  
While in the case of Saurer~1 the derived temperature
yields a reddening consistent with the photometric one, in the case of
Berkeley~29 the spectroscopic reddening turns out to be E(B$-$V)=
0.08, significantly smaller than the photometric one (see the discussion
in Sect~6).\\
\noindent
Initial guesses for the gravity $\log$(g) were derived from:

\begin{equation}
log(\frac{g}{g_{\odot}}) = log(\frac{M}{M_{\odot}}) + 4 \times
log(\frac{T_{eff}}{T_{\odot}}) - log(\frac{L}{L_{\odot}})
\end{equation}

\noindent
taken from \citet{carre97}.  In this equation the mass
$\frac{M}{M_{\odot}}$ was derived from the comparison between the
position of the star in the Hertzsprung$-$Russell diagram and the Padova
Isochrones \citep{gir00}. The luminosity $\frac{L}{L_{\odot}}$
was derived from the the absolute magnitude $M_V$, assuming
the literature distance moduli of 
15.8 for Berkeley~29 \citep{kal94} and 16.0 for Saurer~1 \citep{car03}.
The bolometric
correction (BC) was derived from the relation BC--$T_{eff}$ from
\citet{alo99}.  The input $\log$(g) values were then adjusted in order to
satisfy the ionization equilibrium of Fe~I and Fe~II during the
abundance analysis.  Finally, the micro$-$turbulence velocity is given by
the following relation \citep{gra96}:

\begin{equation}
v_t[km s^{-1}] = 1.19 \times 10^{-3} \times T_{eff} - 0.90 \times log(g) - 2  
\end{equation}

\noindent
The final adopted parameters are listed
in Table~2.

\subsection{Abundance determination}
The LTE abundance program MOOG (freely distributed by Chris Sneden,
University of Texas, Austin) was used to determine the metal
abundances. Model atmospheres were interpolated from the grid of
Kurucz models (1992)  by using the values of
$T_{eff}$ and $\log$(g) determined as explained in the previous section.
During the abundance analysis $T_{eff}$, $\log$(g) and $v_t$ were adjusted
to remove trends in excitation potential, ionization equilibrium and
equivalent width for Fe~I and Fe~II lines.  Table~3 contains the
atomic parameters and equivalent widths for the lines used. The first
column is the name of the element, the second the wavelength in \AA,
the third the excitation potential, the fourth the oscillator strength
$\log$ ({\it gf}), and the remaining ones the equivalent widths of the
lines for the observed stars. \\The derived abundances are listed in
Table~4, together with their uncertainties. The measured iron abundances are
[Fe/H]=-0.38$\pm0.14$ and [Fe/H]=-0.44$\pm0.18$ for Saurer~1 and Berkeley~29, 
respectively. The reported errors are derived from the uncertainties
on the single star abundance determination (see Table~4)
Somewhat 
surprisingly, we do not find extremely
low metal abundances, as one would expect from the large
galactocentric distances and the known abundance gradient
\citep{fri02}.\\

As a check for the entire procedure, we derived the abundances for
the star Arcturus (\citealt{fri03})  following the
same recipe as for the main targets, and obtaining abundance values well
consistent with the values reported by \citealt{fri03}, who
performed the same kind of analysis. The results have been listed in
Tables~4 and 5.

\noindent
Finally, following the method described in \citet{vil04}, we derived
the stellar spectral classification, which we provide in Table~1.

\section{REVISION OF CLUSTER PROPERTIES}
Our study is the first to provide spectral abundance determinations of
stars in Berkeley~29 and Saurer~1. Here we briefly discuss the
revision of the properties of these two clusters which follow from our
measured chemical abundances.

\subsection{Saurer 1}
Saurer~1 was recently studied by \citet{car03}, who, on the basis of
deep VI photometry and a comparison with stellar models, derived and
age of 5 Gyr, a distance of 19.2 kpc and a reddening E(V$-$I)=0.18.
These authors suggest that the probable metal content of the cluster
is around Z=0.008. Here we obtained [Fe/H]=$-$0.38$\pm$0.14, which
corresponds to Z=0.007, thus confirming the photometric estimate. By
using the new metallicity, then, the cluster properties are not
modified in a significant way. We confirm both the photometric
reddening and the Galactocentric distance derived by \citet{car03}.

\subsection{Berkeley 29}
Berkeley~29 was studied by \citet{kal94}, who suggested a reddening
E(B$-$V)=0.21 and a very low metallicity of about [Fe/H]=$-$1.0. By
assuming these values, \citet{kal94} derived a Galactocentric distance
of 18.7 kpc.  The present study provides different results.  In fact,
the spectroscopic reddening turns out to be E(B$-$V)=0.08,
significantly lower than the Kaluzny estimate, and the metallicity
[Fe/H]=$-$0.44$\pm$0.18, significantly higher.  These new values of
reddening and metallicity yield new age and galactocentric distance
estimates, 4.5 Gyr and 21.6 kpc, respectively.  
These estimates have been obtained by comparing the cluster
CMD with the exact  metallicity isochrones  from \citep{gir00}.
Therefore Berkeley~29
is the most distant open cluster in the Milky Way currently known.

\section{THE RADIAL ABUNDANCE GRADIENT}
In Fig.~4 we plot the radial abundance gradient as derived from
\citet{fri02}, which is at present  the most updated  version
of the gradient itself.\\
The clusters included in their work (open squares)
define an overall slope of $-$0.06$\pm$0.01 dex kpc$^{-1}$ (solid
line).
For the sake of the simplicity, 
we assume for the moment that the two new clusters analyzed 
in the present work $-$Saurer~1 and Berkeley~29$-$ are two genuine
Galactic disk old OCs.
They (filled circles) clearly deviate from the general trend.  
However, if we take them  into
account, the slope of the gradient would be much flatter, i.e.
$-$0.03$\pm$0.01 dex kpc$^{-1}$.\\ 
The same conclusion could  be extended
to the age$-$dependent radial abundance gradient. 
\citet{fri02} used 39
clusters to show that the radial abundance gradient becomes flatter
with time (see their Fig.~3), although one has to note that different
age gradients have different spatial baselines and therefore describe
different zones of the disk.  This result contrasts with the previous
analysis by \citet{car98}, where no trend was found with age in a
sample of similar size (37 clusters), except for a slight steepening
of the gradient at intermediate ages (see 
their Fig. 7).  When adding the two new
clusters, Berkeley~29 and Saurer~1, the radial gradient in the older
bin becomes almost completely flat.  
In addition we note that if one adds the
new [Fe/H] determinations for NGC~2506, IC~4651
and NGC~6134 by \citet{carre04}, the gradient disappears
almost completely 
also in the younger age bin, thus confirming the overall trend proposed by
\citet{car98}.\\

\section{ABUNDANCE RATIOS}
\citet{fri03} discussed the trends of abundance ratios for old open
clusters and concluded that all the old OCs for which
abundance ratios are available show scaled solar abundance ratios,
with no correlation with overall cluster [Fe/H] or with
cluster age. The only deviation from this conclusion was found in
Collinder~261, a 8 Gyr cluster, where Si and Na abundances are
slightly enhanced.  The same trend for these two elements is reported
by \citet{bra01} for NGC~6819, a 3 Gyr open cluster with higher than
solar metal abundance.  Apart from these two elements, however, old
OCs show scaled solar abundances, in agreement with typical
Galactic disk field stars.\\

In Table~5 we list the abundance ratios for the observed stars in
Berkeley~29 and Saurer~1. Our program clusters have ages around 5 Gyrs
and iron metal content [Fe/H] $\approx$ $-$0.4. They are therefore
easily comparable with clusters having similar properties discussed
earlier in the literature, such as NGC~2243 and Melotte~66 (see
\citealt{fri03}, Tab.~7).  It is interesting to note that while the
latter two clusters have  scaled solar abundances, our program
clusters have generally enhanced values for all the abundance
ratios.\\

At a similar [Fe/H], Berkeley~29 and NGC~2243 have different values for
all the abundance ratios, with the exception of [Mg/Fe], although the
average [$\alpha$/Fe] ratio is enhanced in Berkeley~29 and solar
scaled in NGC~2243.  Berkeley~29 also exhibits enhanced [O/Fe] with
respect to NGC~2243.\\

The same conclusions can be drawn for Saurer~1, when compared with
Melotte~66, an old open cluster of similar age and [Fe/H]. All the
abundance ratios in Saurer~1 appear enhanced when compared with
Melotte~66, except for [O/Fe], which is similar in the two clusters.

\section{DISCUSSION}
\noindent
By looking at Fig.~4 one can argue that fitting with a straight line
all the data points does not have any  statistical
meaning. Simply, the two new points (filled  circles) deviate
from the bulk of the other points.\\
These results seem instead to imply that
the chemical properties of the outer galactic disk are different from
the bulk of the disk itself. We propose here two possible
explanations:

1) In a series of papers (\citealt{cra03, frin03,roc03}), it has been
shown that a ring$-$like stellar structure, the Monoceros stream, exists
in the direction of the Galactic anticenter, which could represent the
debris of a disrupting dwarf galaxy in a noncircular, prograde orbit
around the Milky Way. In particular, \citet{frin03} proposed that a
number of globular and old OCs might be associated with the
stream. With a radial velocity of 24.6$\pm$0.1 km s$^{-1}$  and a galactic
longitude of 197\fdg98, Berkeley 29 seems to be well associated with
the stream (see Fig.~2 in \citealt{frin03}).  Moreover, its
metallicity of [Fe/H]=$-$0.44$\pm$0.18 perfectly matches the global metal
abundance of the stream ($-$0.4$\pm0.3$).\\ We cannot, however, draw
the same conclusion for Saurer~1. Although the metal abundance is the
right one, the longitude 214\fdg31 and the radial velocity of
104.6$\pm$0.1 km s$^{-1}$ completely rule out the possibility that this
cluster is associated with the stream. One should invoke another
explanation to justify the particular location of this cluster in the
Galactic disk radial abundance gradient.\\

2) Some Galactic chemical evolution models predict that the chemical
evolution of the outer disk is dominated by the halo, and therefore it
undergoes a completely different evolution from the bulk of the disk.
We took the radial abundance gradient predicted by the {\it C-}model
of \citet{chi01}, which assumes that the Milky Way formed through two
main collapse events, the first one generating the halo and the bulge,
the second one smoothly building up the disk inside$-$out.  In this
model,  a higher initial
enrichment due to star formation in the halo is reached in the outer
thin disk.  This is due to
the fact that, although halo and disk formed almost independently, the
halo gas is predominating in the outermost disk regions and it creates
a pre$-$enriched disk gas dominating the subsequently accreted
primordial one. This gives origin to the rise in [Fe/H] at large
galactocentric distances (see Fig.~4, dotted line).  At smaller
galactocentric distances, the amount of disk gas always predominates,
thus diluting the enriched halo gas which falls onto it.  
We finally recall that the negative
abundance gradient predicted for galactocentric distances smaller than
15$-$16 kpc is due, in this framework, to the assumption of an
{\it inside$-$out} formation of the galactic thin disk.  In Fig~4 we show
the expected behaviour of the gradient in the outskirts of the disk,
and an inversion of the slope is predicted to occur at 13$-$14 kpc from
the Galactic center, roughly as observed.
\noindent
We stress that chemical evolution models simply rely on abundances and
abundance ratios, and cannot take into account the kinematics of the
stellar population.  However, the agreement between model and
observations is rather striking. Further support to this scenario
comes from the analysis of the expected abundance ratios in the
outskirts of the disk. Here the models by \citet{chi01} predict
overabundances in O and $\alpha$ elements similar to what we actually
observe, due to the fact that the model evolution is dominated by the old
stellar population of the Galactic halo.

\noindent
It is true that our results rely on four stars in two clusters,
and that more data are needed to clarify the global issue of the evolution
of the outer Galactic disk. However it is quite clear from the present analysis
the the outer disk does not reflect the general trend seen in the inner
disk, and that the outer disk is  probably the outcome a series of mergers
occured in the past, in the same fashion as the Galactic halo.

\section{CONCLUSIONS}
We have presented high resolution spectroscopy of four giant stars in
the extremely distant old OCs Saurer~1 and Berkeley~29, and
provided the first estimate of their metal abundances.  We have found
that these two clusters do not belong to the old open cluster family
of the Milky Way, since they significantly deviate from the global
radial abundance gradient and exhibit abundance ratios typical of the
Galactic halo rather than the Galactic disk.  Moreover they are located
quite high onto the Galactic plane, a position which better fits into
the thick disk of the Galaxy.\\
Berkeley~29 is
presumably associated with the Monoceros stream, whereas Saurer~1 does
not belong at all to the stream, and shows quite different kinematic
features.  Without taking the kinematics into account, the anomalous
behaviour of these two clusters can be explained by assuming that the
outer parts of the disk evolved in a different way with respect to the
bulk of the disk itself, and that the chemical evolution driver at
those distances was the Galactic halo \citep{chi01}. In fact the
abundance ratios, in particular [O/Fe] and [$\alpha$/Fe], show the
typical enhancements expected in an old stellar population like the
Milky Way halo.\\

\acknowledgments
G. Carraro expresses his profound gratitude to Maria Sofia
Randich for crucial help and to Yazan Momany for many useful discussions.

\clearpage



\clearpage


\begin{figure}
\plottwo{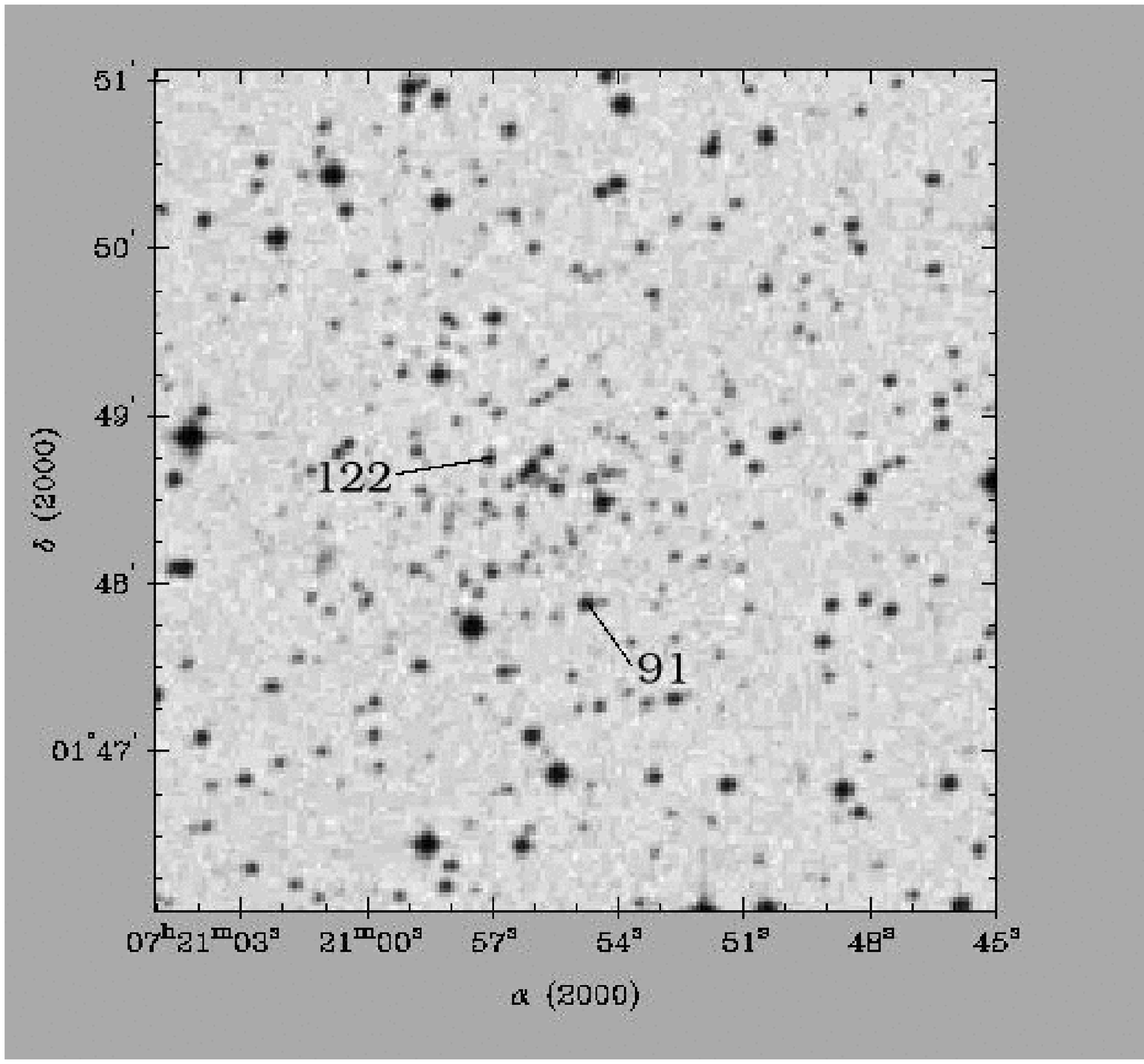}{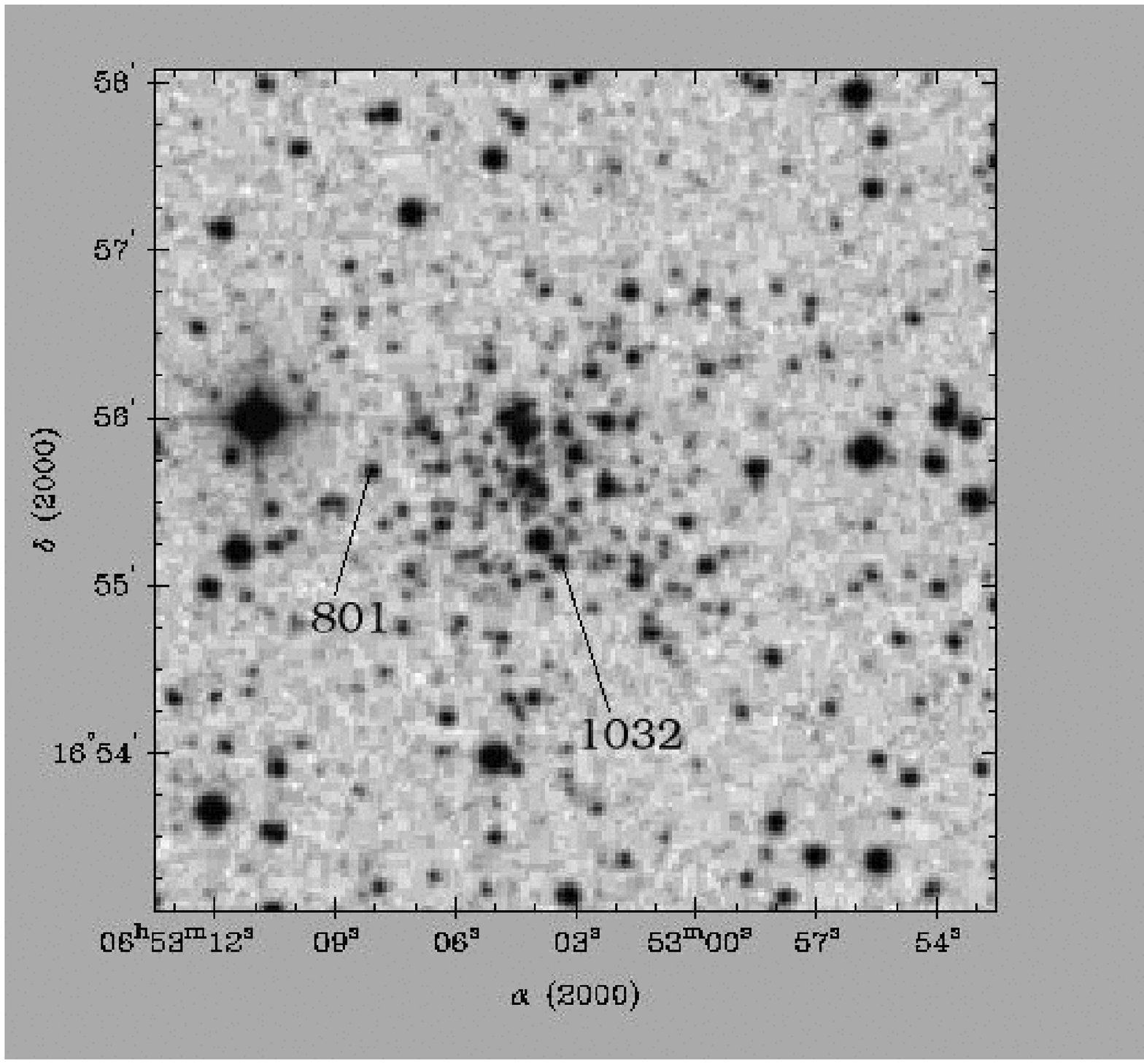}
\caption{Digital Sky Survey (DSS) finding charts of the observed stars in Berkeley~29 (right panel)
and Saurer~1 (left panel)}
\end{figure}


\begin{figure}
\includegraphics[]{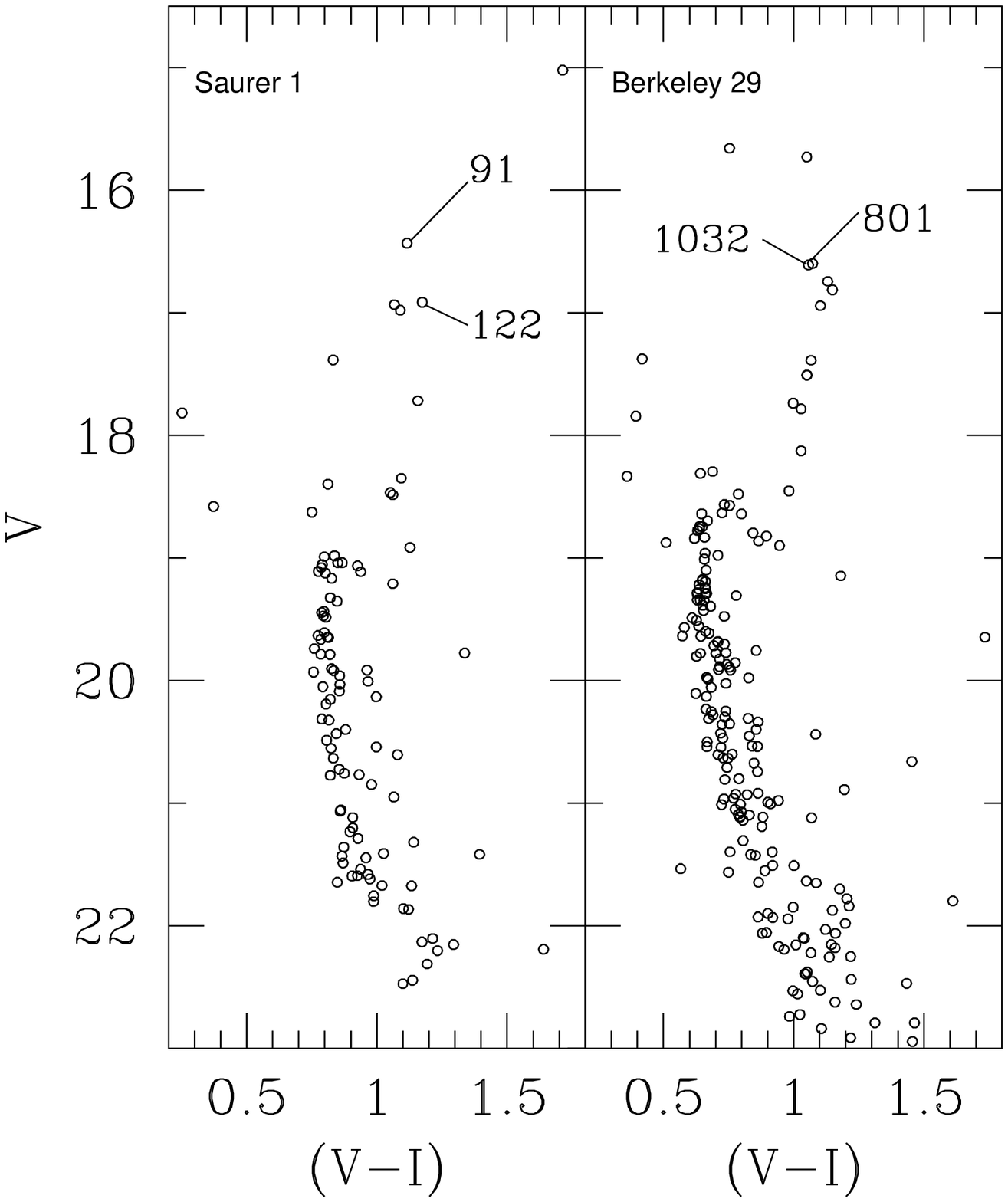}
\caption{Position of the observed stars in the CMD of Saurer~1 (left panel, photometry from  \citealt{car03})
and Berkeley~29 (right panel, photometry from \citealt{kal94}).}
\end{figure}

\begin{figure}
\includegraphics[]{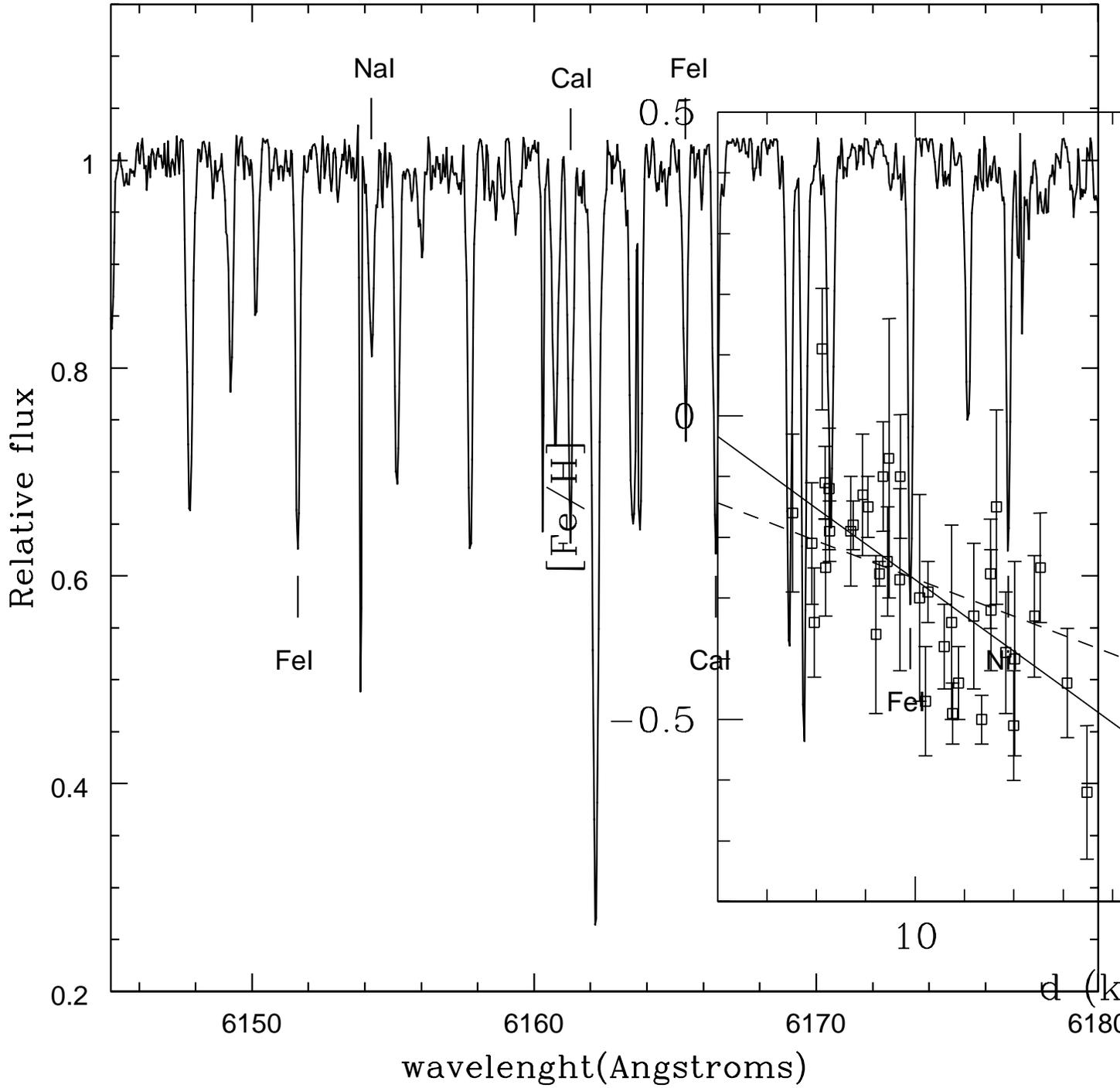}
\caption{An example of extracted spectrum for the star Be~29-801, 
with the main lines indicated}
\end{figure}

\begin{figure}
\includegraphics[]{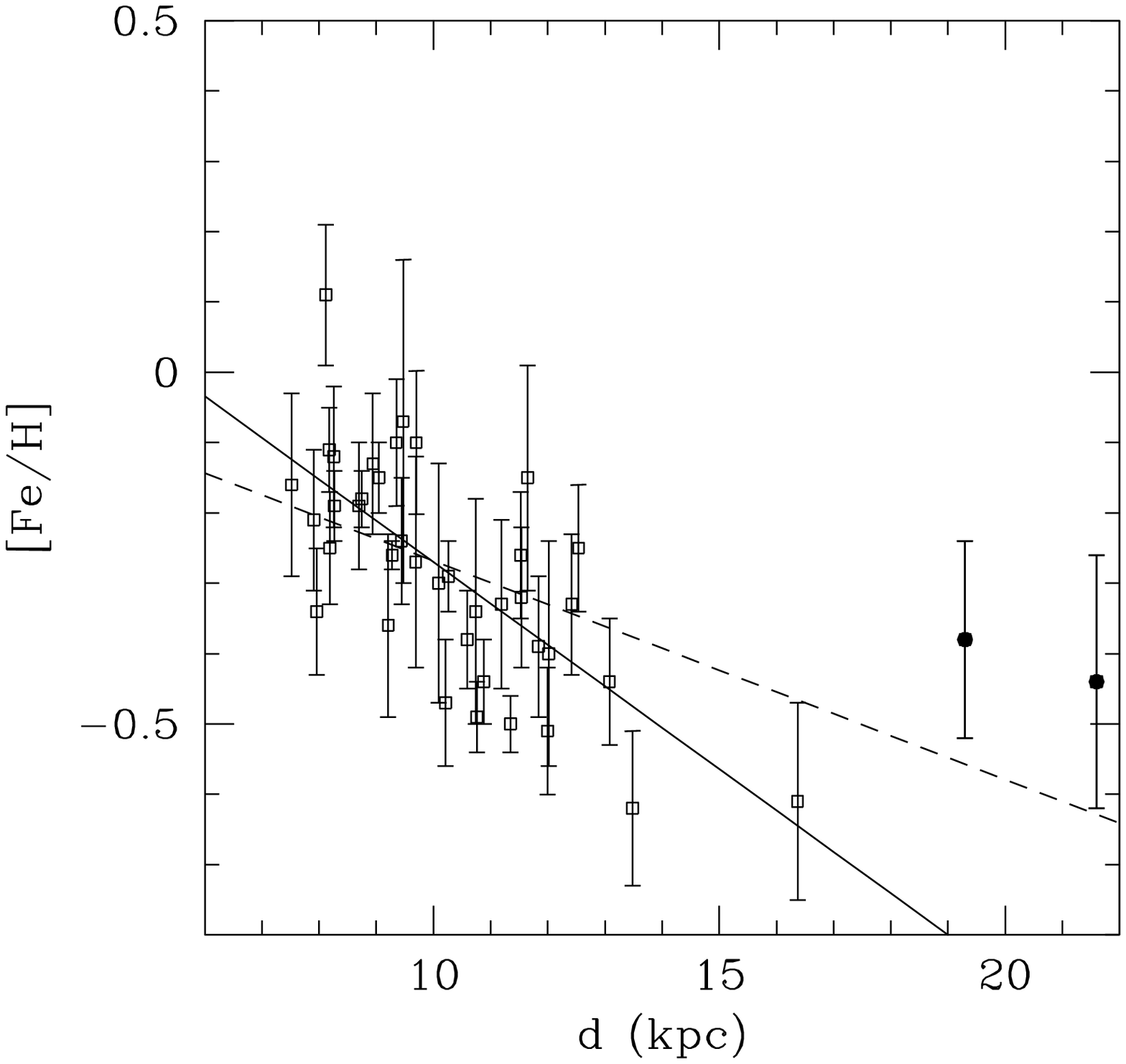}
\caption{The Galactic disk chemical abundance radial gradient. Open
quares are data from \citet{fri02}, whereas filled circles are
Berkeley~29 and Saurer~1 (this work). The solid line is the linear fit to the
\citet{fri02} data, whereas the dashed line is a linear fit 
to all the data points. The dotted line is the radial gradient behavior
expected from a chemical evolution model (\citealt{chi01}).}
\end{figure}






\clearpage

\begin{deluxetable}{lcccccccc}
\tabletypesize{\scriptsize}
\tablewidth{0pt}
\tablecaption{OBSERVED STARS}
\tablehead{
\colhead{ID} & \colhead{RA} & \colhead{DEC} & \colhead{V} &
\colhead{(V$-$I)} & \colhead{$V_{rad}$ (km s$^{-1}$)} &
\colhead{$S/N$} & \colhead{Spectral Type} & \colhead {comments}}
\startdata
Sau 1-91   & 07:20:54.75 & +01:47:53.09 & 16.43 & 1.11    & 
 +104.4$\pm$0.1 &  80 & G7III  & \citet{car03}\\
Sau 1-122  & 07:20:57.08 & +01:48:44.97 & 16.92 & 1.17    & 
 +104.8$\pm$0.1 &  80 & G9III  & \citet{car03}\\
Be 29-801  & 06:53:08.07 & +16:55:40.53 & 16.58 & 1.06    & 
  +24.5$\pm$0.1 &  70 & G6III  & \citet{kal94}\\
Be 29-1032 & 06:53:03.50 & +16:55:08.50 & 16.56 & 1.05    & 
  +24.8$\pm$0.1 &  70 & G6III  & \citet{kal94}\\
\enddata
\end{deluxetable}

\begin{deluxetable}{lccc}
\tabletypesize{\scriptsize}
\tablewidth{0pt}
\tablecaption{ADOPTED ATMOSPHERIC PARAMETERS}
\tablehead{
\colhead{ID} & \colhead{T$_{eff(K)}$} & \colhead{log {\it g} (dex)}
& \colhead{$v_t$ (km s$^{-1}$)} }
\startdata
Sau 1-91   & 5070$\pm$50 & 2.8$\pm$0.1 & 1.60 \\   
Sau 1-122  & 4900$\pm$50 & 2.6$\pm$0.1 & 1.50 \\   
Be 29-801  & 5090$\pm$50 & 2.6$\pm$0.1 & 1.70 \\  
Be 29-1032 & 5070$\pm$50 & 2.6$\pm$0.1 & 1.70 \\   
\enddata
\end{deluxetable}

\clearpage

\begin{deluxetable}{lccccccc}
\tabletypesize{\scriptsize}
\tablewidth{0pt}
\tablecaption{EQUIVALENT WIDTHS}
\tablehead{
\colhead{Element} & \colhead{$\lambda (\AA)$} & \colhead{$E.P.$} &
\colhead{$log\;{\it gf}$} & 
\colhead{$Sa1-91$} & \colhead{$Sa1-122$} &  \colhead{$Be29-801$} &
\colhead{$Be29-1032$}}
\startdata
 Fe I & 5753.120 & 4.240   &  -0.92 &  86.7 &  71.5 & 
78.8 &  99.1\\
 Fe I & 5775.081 & 4.220   &  -1.31 &  68.5 &  66.8 & 
62.4 &  67.5\\
 Fe I & 6024.058 & 4.548   &  +0.16 & 105.0 & 117.0 &
109.0 & 106.7 \\
 Fe I & 6042.226 & 4.652   &  -0.89 &  60.9 & \nodata& 
58.9 &  60.5\\
 Fe I & 6082.72  & 2.22    &  -3.65 &  56.9 &  64.7 & 
66.1 &   -\\
 Fe I & 6151.620 & 2.180   &  -3.27 &  74.6 &  94.0 & 
72.0 &  77.2\\
 Fe I & 6165.360 & 4.143   &  -1.58 &  53.0 &  63.5 & 
51.7 &  43.2\\
 Fe I & 6173.340 & 2.220   &  -2.89 &  98.1 & 105.0 & 
80.5 &  84.0\\
 Fe I & 6229.230 & 2.845   &  -3.04 &  45.2 &  74.7 & 
55.5 &  54.2\\
 Fe I & 6246.320 & 3.590   &  -1.00 & 127.1 & 130.9 &
112.7 & 117.3\\
 Fe I & 6344.15  & 2.43    &  -2.97 &  86.9 & 109.4 & 
90.2 &  95.7\\
 Fe I & 6481.880 & 2.280   &  -2.90 & 104.3 & 102.6 & 
86.5 &  90.8\\
 Fe I & 6574.229 & 0.990   &  -5.00 &  44.1 &  79.9 & 
71.8 &  74.3\\
 Fe I & 6609.120 & 2.560   &  -2.44 &  98.8 & \nodata & 
94.1 &  93.7\\
 Fe I & 6703.570 & 2.758   &  -3.08 &  53.9 &  55.6 & 
51.7 &  53.3\\
 Fe I & 6705.103 & 4.607   &  -1.17 &  55.9 &  66.0 & 
45.3 &  44.1\\
 Fe I & 6820.372 & 4.638   &  -1.05 &  55.1 &  48.8 & 
43.2 &  43.1\\
 Fe I & 6839.831 & 2.559   &  -3.48 & \nodata & \nodata
&  51.3 &  57.9\\
 Fe I & 7461.520 & 2.560   &  -3.64 &  52.1 &  60.9 & 
61.4 &  56.4\\
 Fe I & 7568.900 & 4.280   &  -0.93 & 102.7 &  94.4 & 
75.2 &  80.4\\
 Fe I & 7723.200 & 2.280   &  -3.51 &  79.9 &  91.9 & 
61.9 &   -\\
 Fe I & 7807.909 & 4.990   &  -0.66 & \nodata & \nodata
&  51.0 &  57.9\\
 Fe II& 6084.100 & 3.20    &  -3.95 &  35.2 & \nodata & 
33.0 &  27.9\\
 Fe II& 6149.250 & 3.89    &  -2.81 &  44.4 &  45.8 & 
47.0 &  56.2\\
 Fe II& 6247.560 & 3.89    &  -2.65 &  55.2 &  56.9 & 
56.8 &  65.8\\
 Fe II& 6369.463 & 2.89    &  -4.22 & \nodata & \nodata
&  30.9 &   -\\
 Fe II& 6416.930 & 3.89    &  -2.67 &  58.0 &  53.2 & 
51.1 &  36.0\\
 Fe II& 6456.390 & 3.90    &  -2.35 & \nodata &  66.4 & 
71.5 &  79.0\\
 Fe II& 6516.080 & 2.89    &  -3.34 &  72.2 &  72.9 & 
76.2 &  85.5\\
 Al I & 6696.03  & 3.14    &  -1.44 & \nodata & \nodata
&  34.4 &   -\\
 Al I & 6698.67  & 3.13    &  -1.94 &  32.4 &  35.5 & 
28.9 &  24.0\\
 Ca I & 6161.30  & 2.52    &  -1.07 &  83.0 &  92.1 & 
72.3 &  75.3\\
 Ca I & 6166.44  & 2.52    &  -1.17 & \nodata &  97.4 & 
80.9 &  70.9\\
 Ca I & 6455.60  & 2.52    &  -1.44 &  72.5 &  81.5 & 
72.1 &  58.0\\
 Ca I & 6499.65  & 2.52    &  -0.93 & 107.3 & 119.9 &
101.2 & 103.2\\
 Mg I & 7387.70  & 5.75    &  -1.00 &  62.8 &  69.5 & 
58.7 &  50.0\\
 Na I & 6154.23  & 2.10    &  -1.80 &  50.1 &  48.0 & 
42.8 &  32.2\\
 Na I & 6160.75  & 2.10    &  -1.58 &  73.2 &  73.5 & 
64.0 &  59.9\\
 Ni I & 6175.37  & 4.09    &  -0.68 &  57.6 &  74.9 & 
60.2 &  48.0\\
 Ni I & 6176.81  & 4.09    &  -0.22 &  83.8 &  86.1 & 
72.5 &  68.4\\
 Ni I & 6223.99  & 4.10    &  -1.11 & \nodata &  42.1 & 
33.0 &  35.0\\
 Si I & 5793.08  & 4.93    &  -2.11 & \nodata &  48.1 & 
48.2 &  37.5\\
 Si I & 6142.49  & 5.62    &  -1.86 &  37.6 & \nodata & 
14.1 &  14.0\\
 Si I & 6145.02  & 5.61    &  -1.56 & \nodata &  45.4 & 
34.7 &  39.7\\
 Si I & 6243.82  & 5.61    &  -1.48 &  54.5 &  50.6 & 
54.3 &  43.3\\
 Si I & 7034.91  & 5.87    &  -0.95 & \nodata & \nodata
&  49.8 &  59.4\\
 Ti I & 5978.54  & 1.87    &  -0.52 &  41.5 &  65.4 & 
36.5 &  39.8\\
 O  I & 7771.95  & 9.14    &  +0.28 &  61.0 &  55.0 & 
38.0 & 45.3\\    
 O  I & 7774.18  & 9.14    &  +0.15 &  51.0 &  37.4 & 
50.6 & 33.9\\
 O  I & 7775.39  & 9.14    &  -0.14 &  40.3 &  20.1 & 
30.3 & 32.1\\
\enddata
\end{deluxetable}

\begin{deluxetable}{lcccccccccc}
\tabletypesize{\scriptsize}
\tablewidth{0pt}
\rotate
\tablecaption{MEAN STELLAR ABUNDANCES}
\tablehead{
\colhead{ID} & \colhead{[FeI/H]} &  \colhead{[FeII/H]} &
\colhead{[AlI/H]} & \colhead{[CaI/H]} &
\colhead{[MgI/H]} &  \colhead{[NaI/H]} & \colhead{[NiI/H]} & 
\colhead{[SiI/H]} & \colhead{[TiI/H]} &
\colhead{[OI/H]}  }  
\startdata
Sau 1-91   & -0.38$\pm$0.14 & -0.41$\pm$0.07 & -0.03$\pm$0.15 & -0.22$\pm$0.15 & -0.37$\pm$0.10 & +0.10$\pm$0.15 & -0.23$\pm$0.10 & +0.10$\pm$0.15 & -0.37$\pm$0.15 & +0.11$\pm$0.15\\
Sau 1-122  & -0.38$\pm$0.15 & -0.41$\pm$0.05 & -0.07$\pm$0.15 &
-0.15$\pm$0.16 & -0.33$\pm$0.10 & -0.01$\pm$0.15 & -0.14$\pm$0.11
& -0.10$\pm$0.14 & -0.14$\pm$0.15 & -0.04$\pm$0.15\\
Be 29-801  & -0.45$\pm$0.16 & -0.53$\pm$0.10 & -0.27$\pm$0.17 &
-0.29$\pm$0.16 & -0.39$\pm$0.15 & +0.01$\pm$0.07 & -0.28$\pm$0.11
& -0.21$\pm$0.18 & -0.43$\pm$0.15 & -0.25$\pm$0.15\\
Be 29-1032 & -0.43$\pm$0.21 & -0.48$\pm$0.25 & -0.20$\pm$0.15 &
-0.38$\pm$0.15 & -0.51$\pm$0.15 & -0.12$\pm$0.16 & -0.38$\pm$0.13
& -0.24$\pm$0.13 & -0.40$\pm$0.15 & -0.27$\pm$0.15\\
Arcturus & -0.51$\pm$0.09 & -0.49$\pm$0.06&-0.14$\pm$0.13&-0.19$\pm$0.09&-0.05$\pm$0.11&-0.27$\pm$0.12&
-0.35$\pm$0.20&-0.16$\pm$0.14&-0.29$\pm$0.15&-0.12$\pm$0.15\\
\enddata
\end{deluxetable}

\begin{deluxetable}{lccccccccc}
\tabletypesize{\scriptsize}
\tablewidth{0pt}
\rotate
\tablecaption{ABUNDANCE RATIOS}
\tablehead{
\colhead{ID} & \colhead{[Fe/H]} &  \colhead{[Ca/Fe]} &
\colhead{[Mg/Fe]} & \colhead{[Si/Fe]} &
\colhead{[Ti/Fe]} &  \colhead{[O/Fe]} & \colhead{[Na/Fe]} & 
\colhead{[Al/Fe]} & \colhead{[Ni/Fe]} }  
\startdata
Sau 1-91   & -0.38  &+0.16 & +0.01 & +0.48 & +0.01 &
+0.49 & +0.48 & +0.35 & +0.15\\ 
Sau 1-122  & -0.38  &+0.23 & +0.05 & +0.28 & +0.24 &
+0.34 & +0.39 & +0.31 & +0.24\\
Be 29-801  & -0.45  &+0.16 & +0.06 & +0.24 & +0.02 &
+0.20 & +0.46 & +0.18 & +0.17\\
Be 29-1032 & -0.43  &+0.05 & -0.08 & +0.19 & +0.03 &
+0.16 & +0.31 & +0.23 & +0.05\\
Arcturus & -0.51 & +0.32&+0.46&+0.25&+0.22&+0.39
&+0.24&+0.27&+0.16\\
\enddata
\end{deluxetable}



\clearpage

\end{document}